\documentclass[rsi, reprint,superscriptaddress]{revtex4-1}
\usepackage{amsmath,graphicx,array,wasysym,subfigure}
\DeclareGraphicsExtensions{.pdf}

\begin{document}

\title{A 30 mK, 13.5 T scanning tunneling microscope with two independent tips}
\author{Anita Roychowdhury}
\affiliation{Laboratory for Physical Sciences, College Park, MD}
\affiliation{Center for Nanophysics and Advanced Materials, Department of Physics, University of Maryland---College Park}
\author{M. A. Gubrud}
\altaffiliation[Current address: ]{Woodrow Wilson School of Public and International Affairs, Princeton University }
\affiliation{Laboratory for Physical Sciences}
\affiliation{Center for Nanophysics and Advanced Materials, Department of Physics, University of Maryland---College Park}
\author{R. Dana}
\affiliation{Laboratory for Physical Sciences}
\author{J. R. Anderson}
\affiliation{Center for Nanophysics and Advanced Materials, Department of Physics, University of Maryland---College Park}
\author{C. J. Lobb}
\affiliation{Center for Nanophysics and Advanced Materials, Department of Physics, University of Maryland---College Park}
\author{F. C. Wellstood}
\affiliation{Center for Nanophysics and Advanced Materials, Department of Physics, University of Maryland---College Park}
\author{M. Dreyer}
\affiliation{Laboratory for Physical Sciences}
\date{\today}

\begin{abstract}
We describe the design, construction, and performance of an ultra-low temperature, high-field scanning tunneling microscope (STM) with two independent tips. The STM is mounted on a dilution refrigerator and operates at a base temperature of 30 mK with magnetic fields of up to 13.5 T.  We focus on the design of the two-tip STM head, as well as the sample transfer mechanism, which allows \textit{in situ} transfer from an ultra high vacuum (UHV) preparation chamber while the STM is at 1.5 K. Other design details such as the vibration isolation and rf-filtered wiring are also described. Their effectiveness is demonstrated via spectral current noise characteristics and the root mean square roughness of atomic resolution images. The high-field capability is shown by the magnetic field dependence of the superconducting gap of $\text{Cu}_x\text{Bi}_2\text{Se}_3$. Finally, we present images and spectroscopy taken with superconducting Nb tips with the refrigerator at 35 mK that indicate that the effective temperature of our tips/sample is approximately 184 mK, corresponding to an energy resolution of 16 $\mu$eV. 
\end{abstract}
\maketitle
\thispagestyle{empty}

\newpage
\section{INTRODUCTION}

Since the invention of the scanning tunneling microscope (STM) \cite{Binnig82}, numerous systems have been built that operate at cryogenic temperatures \cite{Elrod84, Marti87, Fein87, Tessmer94, Pan99, Suderow02, Song10, Assig13}. Cryogenic temparatures not only provide the potential for achieving finer energy resolution, but also open up the possibility of exploring otherwise inaccessible phenomena such as superconductivity and a wide range of solid state quantum effects. In the 1990's a variety of 4 K STMs \cite{Tessmer94, Wildoer94, Meyer96, Wittneven97, Ferris98} were developed and used to perform pioneering studies at the atomic scale. Since then, several groups have successfully implemented STM's on $^3$He refrigerators \cite{Pan99, Kugler00, Heinrich02, Suderow02, Wiebe04, Kamlapure13} and dilution refrigerators \cite{Davidsson92,Suderow02,Moussy01,Upward01,Kambara07,Song10,Assig13, Misra13}. A few groups have also built and tested dual-probe STMs, one of which operates at 4.2 K \cite{Grube01, Okamoto01, Wu06}. Multi-probe STMs that operate in UHV conditions, with cryogenic capabilities in the 4K--10K range, are also commercially available \cite{Oxford,Attocube, RHK, mprobes}. Despite these advances, there are still relatively few instruments that operate reliably at sub-Kelvin temperatures. This is due to the inherent difficulties in constructing a system that operates at ultra-low temperatures, in ultra-high vacuum, with the mechanical stability necessary for STM measurements. 

In this article, we describe the design, construction, and operation of a millikelvin STM system with a magnetic field and UHV sample preparation capabilities. Our microscope possesses two scanning systems that allow us to independently acquire topographic images of a sample at mK temperatures. In the absence of a tip-exchange mechanism, an STM with two independent scanning systems provides a relatively simple way to study different regions of the same sample simultaneously or in quick succession. The instrument was also designed to ultimately allow two superconducting tips to be connected via a flexible superconducting wire forming a SQUID loop with a superconducting sample. In such a modified instrument, one STM tip could provide a reference point against which the gauge invariant phase difference could be mapped out on the sample surface \cite{Sullivan13}.
\begin{figure*}[t]
\centering
\includegraphics[width= \textwidth]{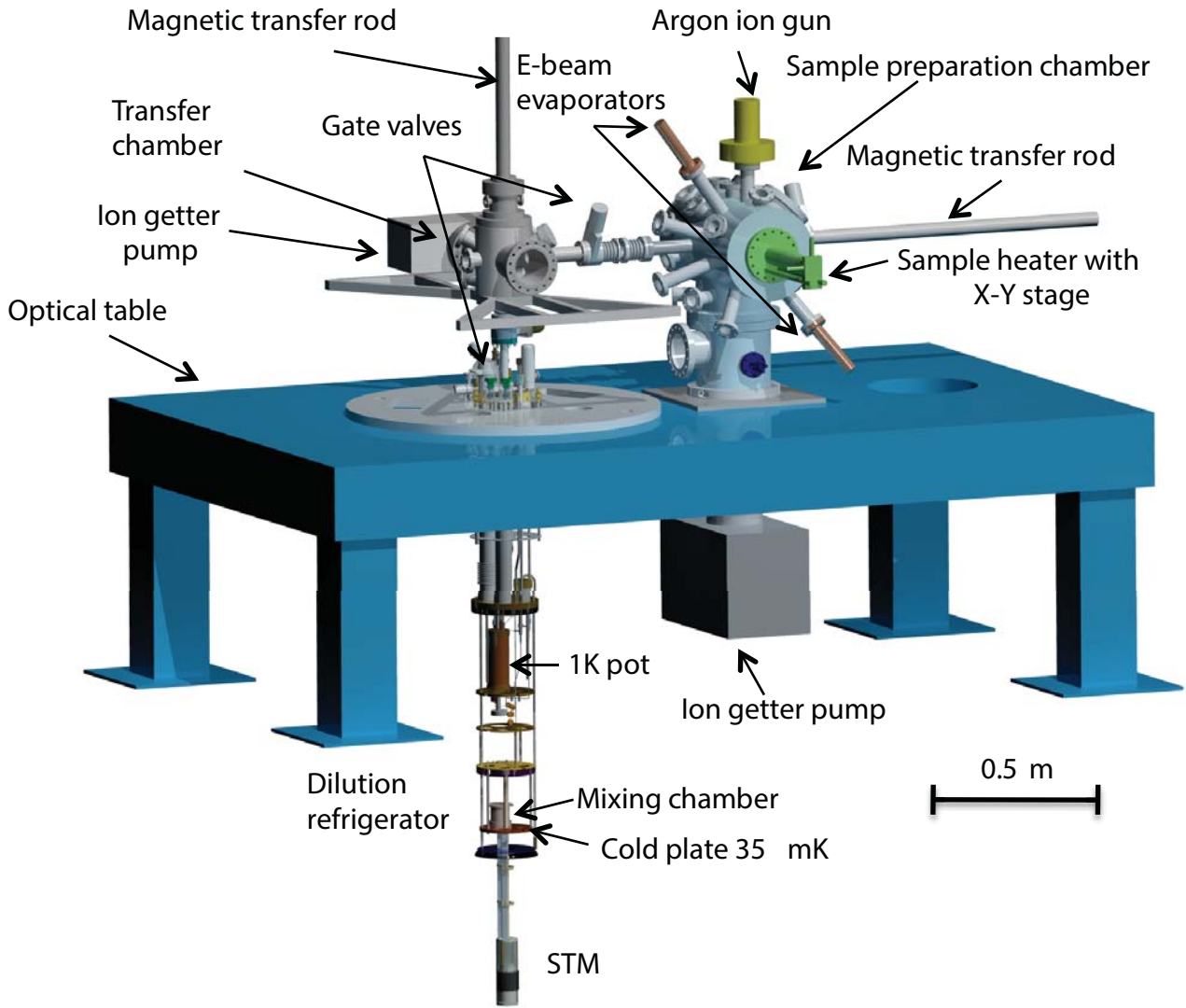}
\caption{(Color Online) Overview of dual tip mK STM system. The dewar and radiation shields that surround the dilution refrigerator are not shown.}
\end{figure*}
\section{SYSTEM DESIGN}

Designing an STM that can operate at millikelvin temperatures involves satisfying several stringent requirements. First, the STM tip and sample need to be cooled to mK temperatures. Second, the sample must be kept in vacuum at a pressure of $10^{-11}$ Torr or better to prevent contamination of the surface. Third, the system must be kept as vibrationally isolated as possible---ideally with the tip-to-sample distance stable to better than 5 pm in order to resolve atomic scale features in topographic images and fine scale structure in spectroscopy. For the same reason, we require voltage resolution of 10 $\mu$V or better, and current resolution as fine as 1 pA. In addition to the above requirements, there are other features that are desirable, such as the ability to prepare and transfer samples in UHV. 

Figure 1 shows an overview of our system. A UHV sample preparation and transfer system sits on an optical table directly above a $^3$He-$^4$He dilution refrigerator that has the STM assembly mounted to the mixing chamber. A vertical sample transfer rod runs along the central axis of the refrigerator to enable top loading of samples from the UHV system into the microscope. The dilution refrigerator is an Oxford Instruments Kelvinox with a cooling power of 400 $\mu$W at 100 mK and a no-load base temperature of 6-7 mK. The operating base temperature for our STM however is 30-35 mK due to a thermal heat-leak of about 30 $\mu$W from the system wiring. Radiation shields clamped to the mixing chamber and the still (0.7 K) prevent black-body radiation from higher stages from reaching the STM and a custom made super-insulated liquid-helium cryostat \footnote{Kadel Engineering, Danville IN.} provides a 4 K bath. This dewar is interchangeable with a custom made 13.5 Tesla superconducting magnet dewar with a LHe capacity of 140 liters \footnote{American Magnetics, Inc.}. The magnet is a high field (vertical bore) solenoid system, with a rated field of 13.8 kG, and a rated current of 99.8 A. The entire setup sits inside a copper and steel walled rf shielded room with a nominal rf attenuation of 100 dB or greater from 1 kHz to 10 GHz.  

\subsection*{The UHV Sample Preparation Chamber}

The UHV subsystem consists of a preparation chamber and a transfer chamber that are separated by a gate valve (see Fig.\ 1). Each chamber has its own ion getter \footnote{Varian Model Starcell 75 and 300, respectively}, titanium sublimation pumps \footnote{Vacuum Generators Model ST2 and Varian Model 929-022}, and ion gauge \footnote{VG Scienta Model VIG18/Arun Microelectronics Ltd. PGC1}. The system is roughed out using a detachable turbo pump \footnote{Varian TModel V551 Navigator/VT 1000HT} backed by an oil-free scroll pump \footnote{Varian Model TriScroll600} and can be baked to $150^\circ$C using heater tape. A load lock allows the introduction of samples into the preparation chamber. Samples are mounted on sample studs that fit into the STM and can be moved throughout the UHV system on a transfer plate using a magnetic transfer rod. 
 
 The sample preparation chamber has a residual gas analyzer \footnote{Extorr Model XT200}, two electron beam evaporators \footnote{Omicron Model EFM3/EVC 100}, and an argon ion sputter gun \footnote{Specs Model IQE 10}. The sample stage in this chamber is attached to an XYZ manipulator \footnote{VG Scienta Model HPT}. Samples may be heated up to $600^\circ$C by a resistive heater or by a direct current heater \cite{Dreyer10}. The preparation chamber also has a room temperature Pan-style STM that operates in UHV. This enables us to examine samples during or after preparation and before we transfer them into the mK STM.

\subsection*{Sample Transfer}
\begin{figure}[tb]
	\includegraphics[height=3.5in]{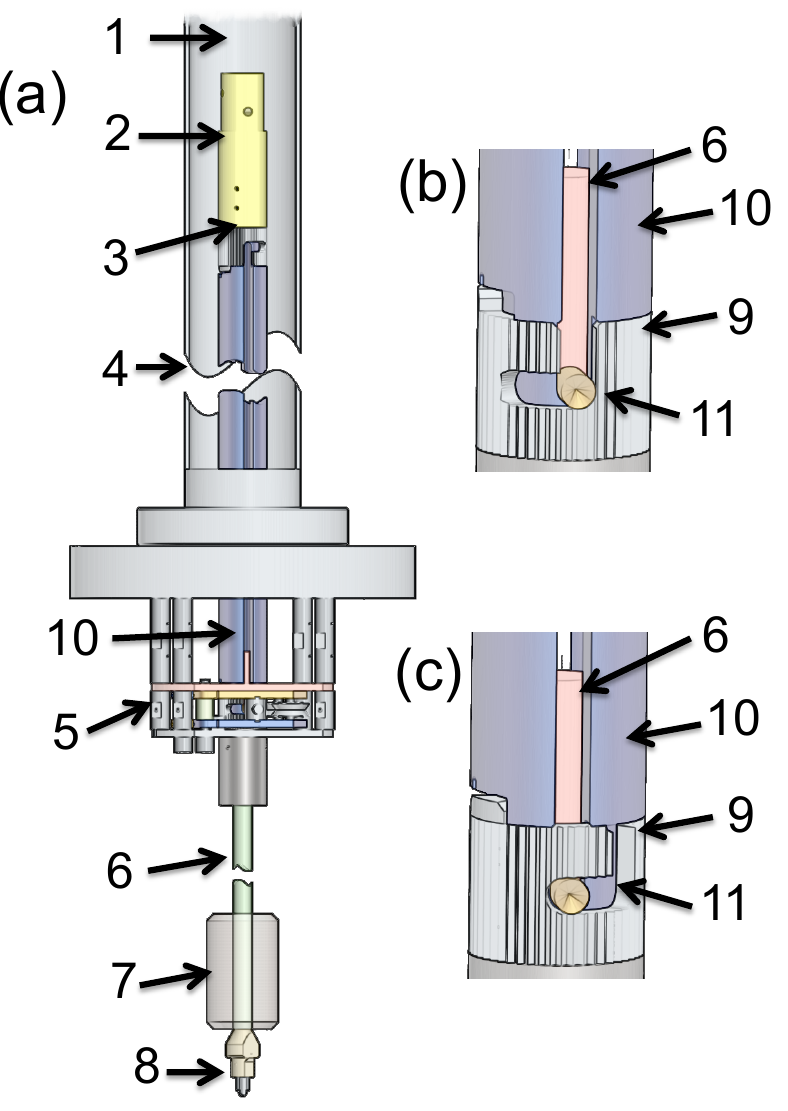}%
	\caption{\label{Fig:CTR_ov}(Color Online) (a) Overview of collapsible transfer rod. (b-c) Locking mechanism in the (b) open and (c) closed position. (1) Shell of magnetic transfer rod, (2) coupler, (3) upper lock, (4) outer rod, (5) clamp with lower lock inside, (6) inner rod, (7) guide piece, (8) sample grabber, (9) lock, (10) outer shaft, (11) slot in outer shaft showing brass pin mounted on inner rod.}
\end{figure}
\begin{figure}[tb]
	\includegraphics[height=1.6in]{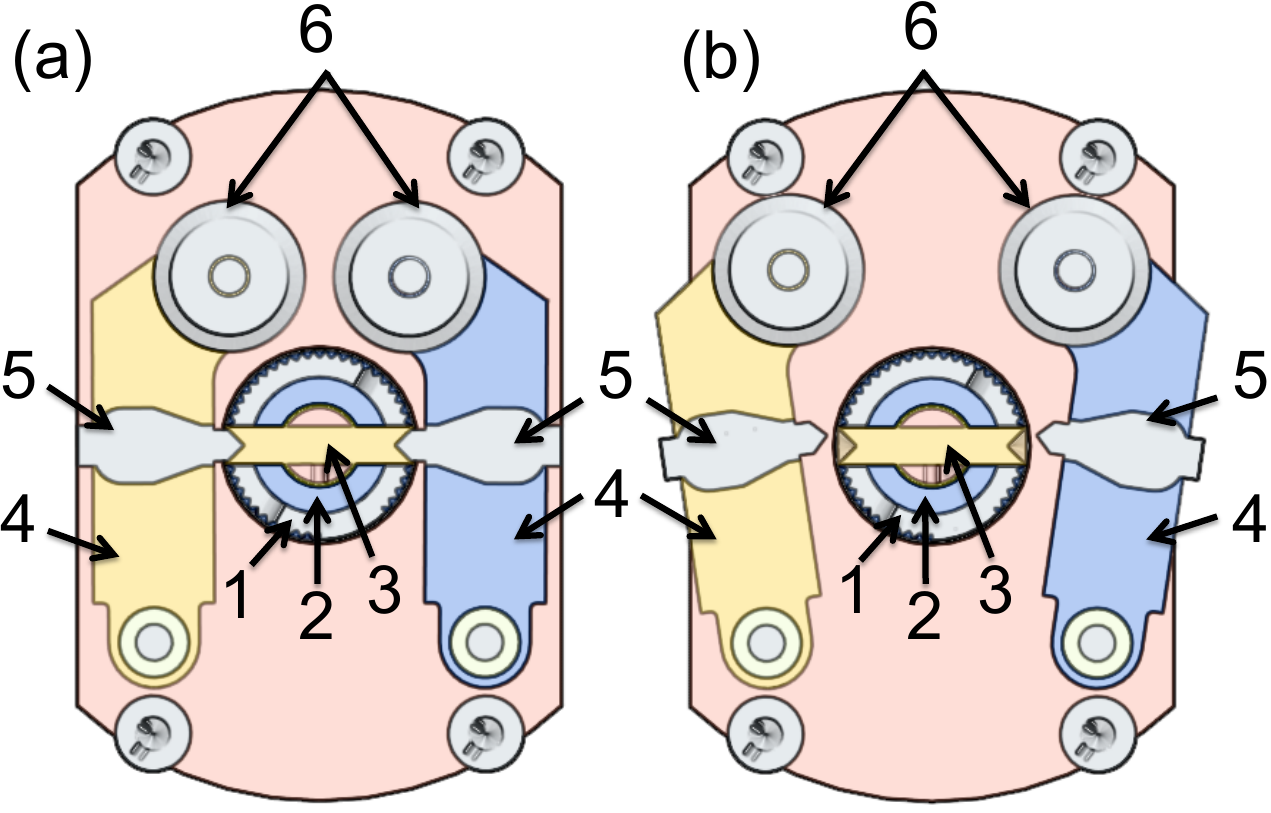} \hspace{1cm}%
	\caption{\label{Fig:CTR_clamp}(Color Online) Transfer rod clamping mechanism in the (a) open and (b) closed position. (1) lock, (2) outer shaft, (3) brass pin mounted on inner rod, (4) clamp arm, (5) spike and (6) pulley system.}
\end{figure}

As we noted above, a very desirable feature in a low-temperature STM is the ability to exchange samples without having to warm the system to room temperature. Our design allows for top loading of samples via a 3 cm diameter access shaft in the  dilution refrigerator. This top-loading system uses a collapsible rod that fits inside a 4' long magnetic transfer rod [Fig.\ 2(a)]. The collapsible rod has an outer tube made from a 0.5" diameter thick-walled aluminum tube with a slot cut along almost the entire length. Inside the slot there is a brass pin that is attached to a thin-walled stainless steel tube. The lower end of the inner tube is fitted with a sample exchange fork \cite{Dreyer10} for engaging the sample stud and a guide piece. The brass pin can be locked in place at either end of the outer tube. The locks have an L-shaped cut [Figs.\ 2(b) and (c)] and are operated by a short wobble stick while the pin is held in place by a clamping mechanism. Figure 3 shows the two spring-loaded clamp arms. The springs force the clamping arms open while a pulley system is used to close them. The clamps are operated by a push/pull feedthrough hooked into a steel rope loop running around the pulleys. Each arm has a short spike. The spikes fit into counter bores on the brass pins in the inner stainless steel tube. This arrangement allows the inner rod to be held in place while the outer rod is moved between the upper and lower latching positions, allowing transfer from of the sample from the transfer chamber to the mixing chamber. 

\subsection*{Wiring and Filtering}

The two main kinds of wires in our system are (1) signal wires for the two STM tips, sample, and capacitive sensors, and (2) wires for the piezo drives and thermometry \cite{gubrud10}.

\begin{table*}[t]
\caption{Wiring for dual-tip mK-STM}
\begin{tabular}{| >{\begin{center}}m{.9in}<{\end{center}} | >{\begin{center}}m{.9in}<{\end{center}} | >{\begin{center}}m{.9in}<{\end{center}} | >{\begin{center}}m{.9in}<{\end{center}} | >{\begin{center}}m{.9in}<{\end{center}} | >{\begin{center}}m{.9in}<{\end{center}} | >{\begin{center}}m{.9in}<{\end{center}} |} \hline
Temperature \\ range & Wire type and Length [m] & Dimensions \\ O.D. [mm] & Resistance \\ \ [$\Omega$/m] & Capacitance \\ \ [pF/m] & 10 GHz Atten. \\ \ [dB/m] & Number \tabularnewline \hline\hline
300 K -- 10 K \\ 10 K -- MXC & CuNi coax\textsuperscript{a} \\ $\sim2$ m & CuNi 0.08 \\ PTFE 0.26 \\ CuNi 0.40 & wire 75 \\ shield 5.2 & 96.2 & 61 & 12 \tabularnewline \hline
300 K -- 10 K \\ 10 K -- 1.4 K & Constantan loom\textsuperscript{b} \\(twisted pairs) \\ $\sim3.5$ m & Constantan 0.10 \\ polyester 0.12 & 66 & $\sim$50  & $\sim$100 \\  & 48  \tabularnewline \hline
1.4 K -- MXC & CuNi-clad NbTi loom\textsuperscript{c} \\ (twisted pairs) \\$\sim1$ m & NbTi 0.05 \\ CuNi 0.08 \\ polyester 0.10 & 52 & $\sim$50  & $\sim$100 \\  & 48  \tabularnewline \hline
MXC-STM & Cu coax\textsuperscript{d}  \\$\sim 0.7$ m & Cu 0.29 \\ PTFE 0.94 \\ Cu 1.19 & wire \\ 0.26 & 96.1 & 3.7 & 48 \tabularnewline \hline
\end{tabular}
\textsuperscript{a}SC-040/50-CN-CN Coax Co., Lt., Kanagawa, Japan.  \textsuperscript{b}Brittania House, Cambridge, UK.  \textsuperscript{c}Brittania House, Cambridge, UK.  \textsuperscript{d}Microstock Inc., Westpoint, PA, USA.
\end{table*}

To achieve good energy resolution, the signal wires must be shielded and filtered to prevent heating and smearing of non-linear electrical characteristics by broad-band noise. The length of each wire (see Table 1) was chosen to reduce heat leaks between stages. Since the body of our STM is made from Macor, which is thermally quite insulating, the wires from the STM are used to carry away heat generated during coarse approach and tunneling measurements. The wire insulation must work with tunneling resistances that range up to 100 G$\Omega$, and must be able to withstand several hundreds of volts applied during field emission in order to clean the tip.  

One potential source of noise in our system is microphonic pickup on the signal wires for the tips. To reduce this pickup we used semi-rigid coax for all signal wires inside the vacuum space of the cryostat. This also provides good shielding against rf pickup and allows us to use rf-tight connectors. To minimize thermal load, we used CuNi microcoax (see Table 1). The coax is clamped to 3.35 cm long Cu posts at the 600 mK still stage, and 7.5 cm long posts at the 100 mK  stage. At the mixing chamber, these lines are connected to bronze powder filters \cite{gubrud10} by SSMC connectors. From the mixing chamber filters to the STM, the lines are semi-rigid coax with Cu shielding, and a Ag plated Cu inner conductor to allow heat to be conducted away from the STM tip, sample, and piezos. 

Four woven wiring looms \footnote{Brittania House, Cambridge, UK}, each with 12 twisted pairs of $\approx 100 \ \mu$m diameter wire were used for the thermometry and piezo wiring. From 300 K to 1.4 K, two Constantan wiring looms were used as they provided a low heat load and resistances that varied only slightly with temperature. From 1.4 K to the mixing chamber, we used two CuNi-clad NbTi looms that were superconducting below 9 K. These lines were connected to the STM via semi-rigid coax as was done for the signal wires. The looms are wrapped around copper heat sinks at each stage on the refrigerator. To ensure adequate thermal anchoring, 40 mm tall heat-sink posts were used at the 4 K, 1.4 K, and 0.7 K stages, and 75 mm tall posts were used at the 100 mK shield stage and mixing chamber. 

To prevent external rf interference from reaching the STM, the signal wires are filtered with low pass $\pi$-filters on entry into the shielded room and at the cryostat. The CuNi microcoax used for the signal wires also provides some filtering. The cable length of 1 m from room temperature to 4 K, and 1 m from 4 K to the mixing chamber, provides a total attenuation of about 39 dB at 1 GHz and 120 dB at 10 GHz \footnote{Rami Ceramic Industries, Nazareth, Palestine}. 

The bronze powder filters used on the signal lines were made from 5.0 m of 50 $\mu$m  Cu-clad NbTi wire, wound in four sections around rods made from Stycast 2850 FT, with alternating chirality to minimize magnetic coupling between the filters. These were embedded in a bronze powder/epoxy composite for increased thermal conductivity \cite{Milliken07}. The filters were tested to about 500 V and the attenuation scales approximately as $\sqrt{f}$ with about 35 dB attenuation at 100 MHz \cite{gubrud10}. 

\subsection*{Vibration Isolation}

Vibration isolation is critical to the operation of an STM because the tunnel current $I$ depends exponentially on the tip-sample spacing $z$, {\it i.e.} $I = I_0e^{-z/z_0}$. Typically the current changes by an order of magnitude when the spacing is changed by 0.1 nm which gives $z_0 = 0.23$ nm. A small change $\delta z$ in $z$ leads to a change in current $\delta I = (I/z_o)\delta z$. Hence to stabilize a 1 nA current to 10 pA requires $\delta z \le 2.3$ pm. This illustrates the extreme stability required for the tip/sample spacing. 

We chose a Pan-style design \cite{Pan99} for the STM head because it is resistant to vibrational noise. Vibration isolation of the entire system is achieved by mounting the refrigerator on an optical table with a pneumatically damped air suspension system \footnote{ Technical Manufacturing Corporation.,TMC Model 71-473-spl}. The nominal isolation efficiency is 97$\%$ at 5 Hz and 99$\%$ at 10 Hz, with a resonance frequency of 0.8--1.7 Hz for the vertical mode. The horizontal mode has an isolation efficiency of up to 90$\%$ at 5 Hz and 95$\%$ at 10 Hz with a resonance frequency of 1--2 Hz \footnote{TMC Model 71-473-SPL; Technical Manufacturing Corporation, Peabody, MA.}.  Another potential source of vibrations is whistling in the 1 K pot, as liquid $^4$He is drawn in through the siphon. To eliminate whistling, the pot was custom-made to a larger size, allowing data to be taken for 1.5 days between fillings.  

\begin{figure*}[tb]
 \centering
 \includegraphics[width=  \textwidth]{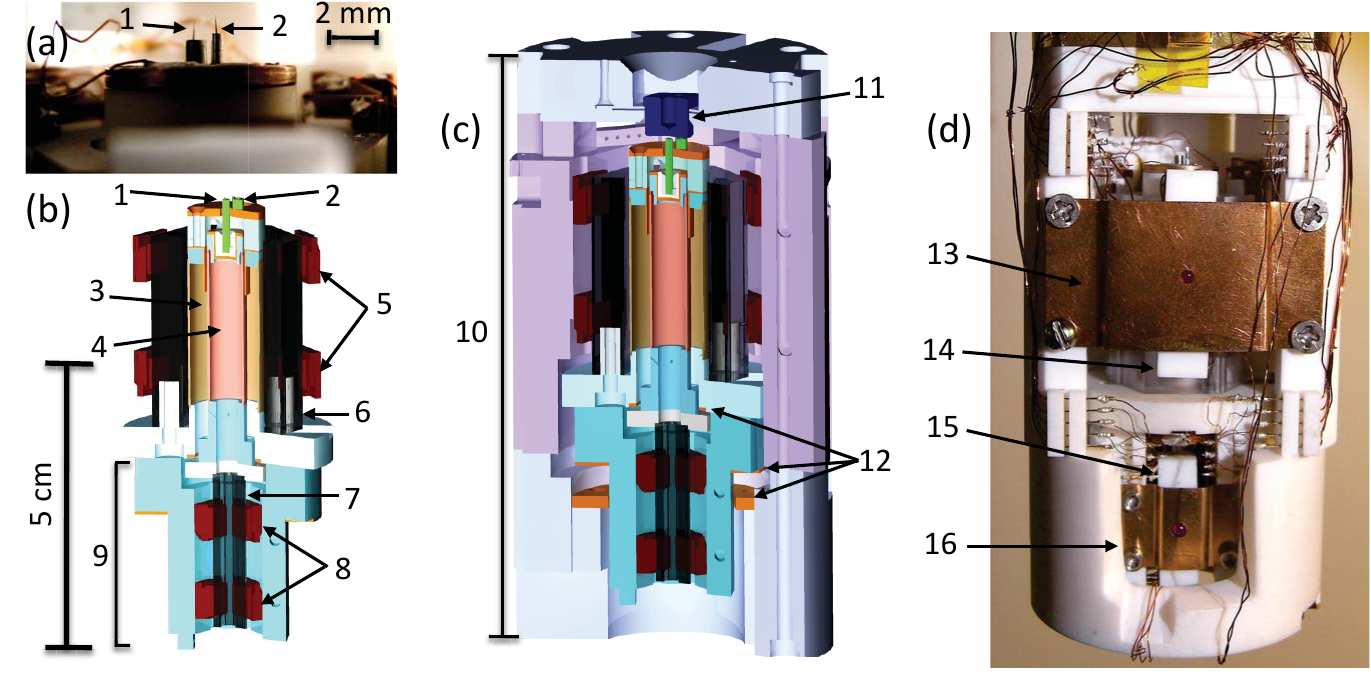}
 \caption{(Color Online) (a) Close-up photograph of STM tips, (b) inner and outer tip assemblies, (c) cross section view and (d) photograph of the STM. The sample stud (11) is inserted from the top, and the tips approach it from below. The inner tip assembly consists of the inner tip (1), attached to the small piezo scanner (4). The outer tip (2) is likewise attached to the large piezo scanner (3). The coarse approach mechanism for the outer tip is formed by a sapphire prism (6) that is driven by six large piezo stacks (5) and (14), four of which are epoxied to the outer MACOR body (10). The coarse approach mechanism for the inner tip (9) rides on the outer coarse approach assembly and has a small sapphire prism (7) that is driven by six small piezo stacks (8) and (15). To set the frictional force between the prisms and the piezo stacks, two piezo stacks for each stages are held against the sapphire prisms by adjustable copper spring plates (13) and (16). The capacitance plates (12) keep track of the relative position of each tip assembly.}
\end{figure*}

To reduce vibrations from the rotary circulation pumps, we place them on large rubber pads in an adjacent room. The pipes from these pumps go through a sand box and vertical and horizontal aluminum bellows for additional vibration isolation.  Pumping line connections to the top of the optical table  present a critical vibration isolation problem. We use bellows in a ``sideways T design'' \cite{gubrud10} to minimize the transfer of pressure variation from the pumping lines to the table. The T-bellows box rests on a rubber pad and has 113 kg of Pb added to it. 
\subsection*{The Dual-Tip STM}

The Macor STM body is bolted rigidly to the mixing chamber plate of the dilution refrigerator via a Cu flange and three L-shaped Cu bars. The lowest internal vibrational modes of the mixing chamber plate are stiffened with the help of additional braces \cite{gubrud10}. The STM body is enclosed in a Au-plated Cu can for electrical and heat shielding. This electrical shield is surrounded by a 100 mK heat shield, which is in turn enclosed by the  inner vacuum can (IVC) and the LHe dewar. 

Our dual-tip STM (see Fig.\ 4) is a modified version of an STM design by Pan {\it et al.} \cite{Pan99}. The main difference is that we have double the number of parts in order to allow the independent operation of two STM tips. Key aspects of Pan's design are its emphasis on mechanical rigidity and the use of materials that are matched as closely as possible to minimize stress from thermal expansion. In our case, this was achieved by making the main STM body from a single 94 mm long Macor cylinder with an outer diameter of 50.8 mm and an inner diameter of 35.6 mm.

Figures 4(b) and 4(c) show details of our STM design. The sample stud (11) is loaded from the top, and the two tips approach the sample from below. The ``inner" tip (1) is attached to the center or ``inner" piezo tube (4), and the ``outer"  (off-center) tip (2) to the ``outer"  piezo tube (3). The tips are side-by-side, about 1 mm apart while the piezo tubes are concentric and share a common axis (see Fig.\ 4(b)). Each piezo tube is epoxied onto a Macor scanner holder.  The large outer scanner (3) operates in a hole in a large sapphire prism (6). The small scanner tube (4) is located above the corresponding small sapphire prism (7) rather than within it. 

The coarse approach mechanism is driven by six shear piezo stacks for each tip (see Fig.\ 4(b)). Four large stacks (5) are epoxied to the inside of the outer Macor STM body (10) and two to a Macor piece (14) held in place by a copper spring plate (13). Likewise, four small stacks (8) are epoxied onto the  smaller inner Macor body, and two to a small Macor piece (15) held in place by a small Cu-Be spring plate (16).  With this arrangement, the outer walker carries the weight of the inner walker. For first approach, the tips are positioned with the inner tip just behind the outer tip (out of tunneling range), and we drive just the large piezo motor so that the outer tip makes contact with the sample first. We then retract the outer tip, and approach with the small piezo motor until we detect tunneling with the inner tip. 

We found that the standard design of four piezos per stack \cite{Pan99} was not strong enough to reliably approach the sample at mK temperatures. Accordingly we use six piezos per stack. Testing of the six-piezo stacks at various voltages through their whole range of motion revealed that they walked about twice as fast as our four-piezo setup. The small walker for the inner tip carries less weight and four piezos per stack are sufficient. We note that the tight space makes wiring the small stacks problematic, and the prototype developed shorts after a couple of years of use. Hence, our current design for the small walker has replaceable piezo stacks epoxied onto copper plates that are screwed onto the small Macor body.
\section{Operation}

Samples are mounted on copper, stainless steel, or molybdenum sample studs that can be changed \textit{in situ}. A sample is either attached to the holder with silver epoxy or  held in place by an L-shaped clamping arm. We use etched W and Nb wires as STM tips \footnote{A. Roychowdhury, et al., ``Plasma Etching of Superconducting Niobium Tips,'' in preparation.}. The tips cannot be changed {\it in situ}---the microscope has to be warmed to room temperature and the vacuum can vented in order to replace a tip. 

After STM tips are installed, the vacuum can is evacuated to  $\sim 10^{-7}$ torr. We then fill the dewar with liquid N$_2$ and bleed $^3$He exchange gas into the inner vacuum can (IVC). Overnight cooling brings the STM to 77 K. The nitrogen is then siphoned out and slowly replaced with liquid helium. After cooling to 4.2 K in about a day, the exchange gas is then pumped out, the 1K pot filled, and standard operation of the dilution refrigerator begins. After reaching a base temperature of 30-35 mK, we insert a sample of single crystal Au(111) or Au(100), and clean each tip using high voltage field emission. After cleaning, the $I$-$V$ curve for each tip is examined to ensure we have a stable, sharp and metallic tip. The gold sample is then exchanged via the sample transfer system for other samples of interest. Sample exchange takes 12-15 hours, including about 3 hours spent pre-cooling the rod at the top plate and about 3 hours pre-cooling at the 1 K stage.

\begin{figure}[t]
 \includegraphics[width= .5\textwidth]{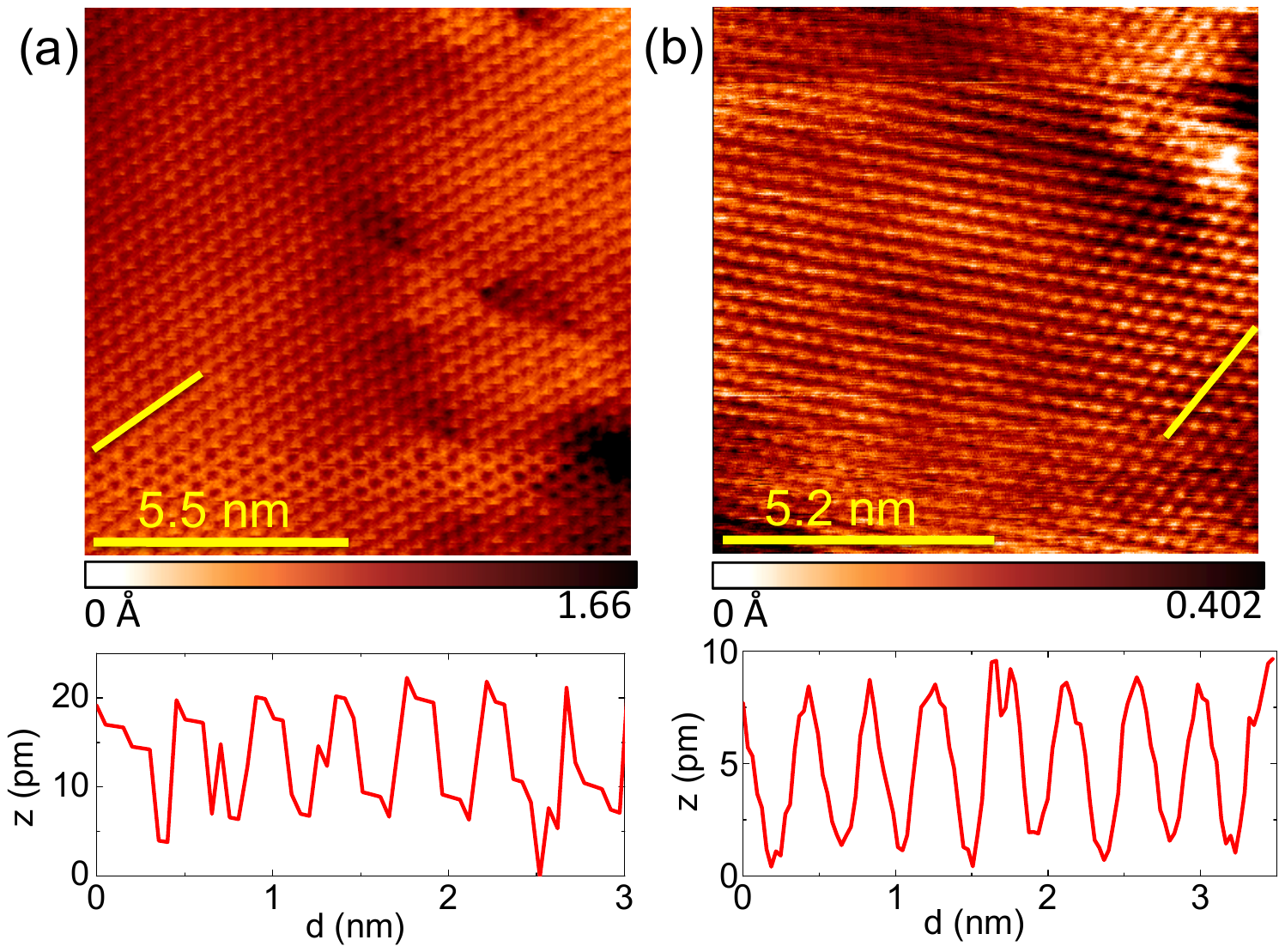}  
\caption{(Color Online) Unfiltered atomic resolution images of $\text{Bi}_2\text{Se}_3$ taken using Nb tips at 35 mK with (a) outer tip and (b) inner tip. Top: Topographic images.  Bottom: Graphs of tip height $z$ vs.\ distance scanned $d$, showing atomic corrugation with peak to valley heights of $\sim$18 pm for the outer tip (a) and $\sim$8 pm for the inner tip (b).}
\end{figure}

\begin{figure}[t]
 \centering
  \includegraphics[width= .48\textwidth]{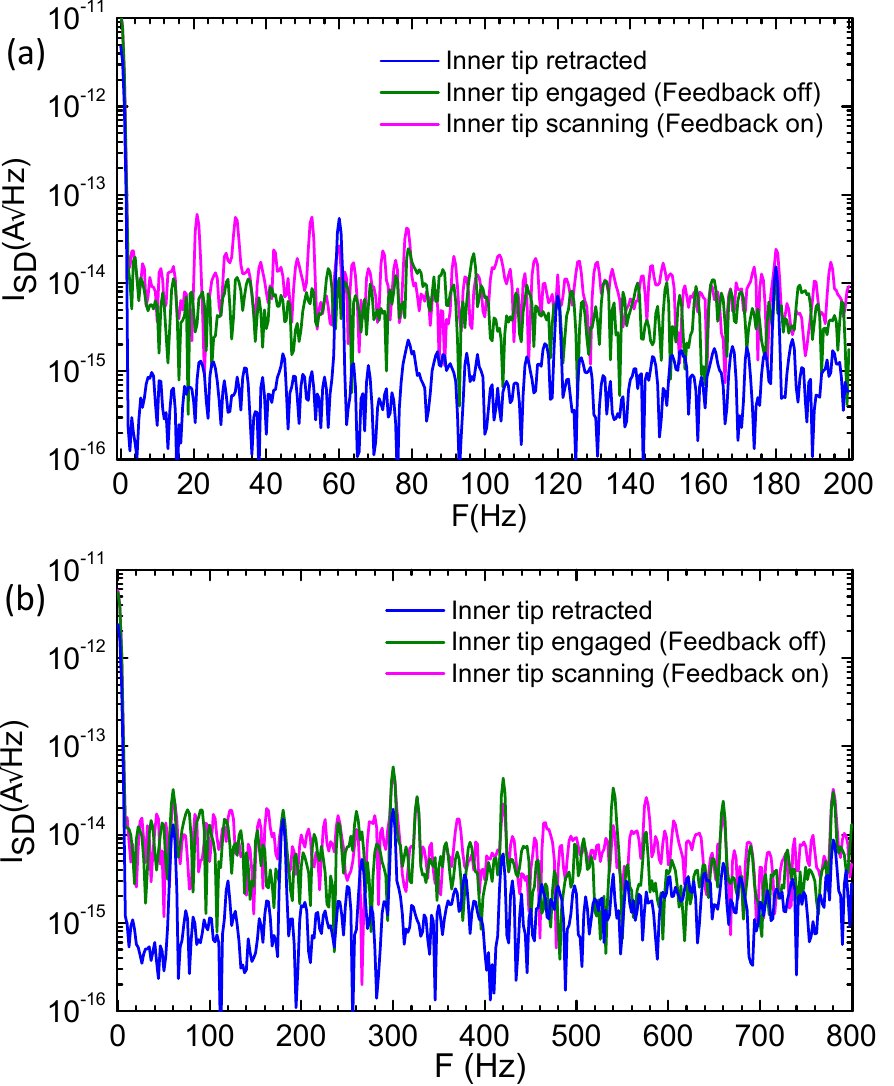}  
\caption{(Color Online) Spectral density of the inner tip tunneling current with the tip retracted (blue), engaged (green), and scanning (pink) for the frequency ranges (a) 0--200 Hz and (b) 0--800 Hz.}
 \end{figure}

\begin{figure}[t]
 \centering
 \includegraphics[width= .49\textwidth]{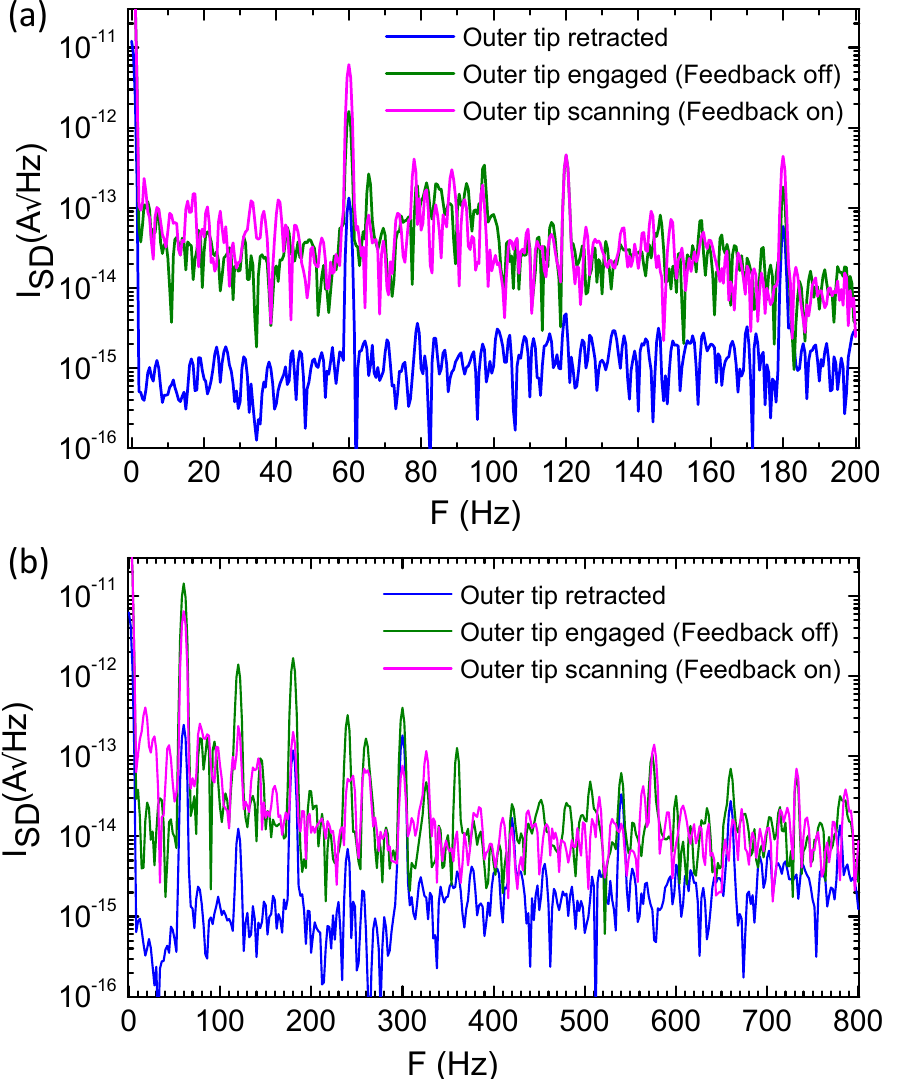}  
\caption{(Color Online) Spectral density of the outer tip tunneling current with the tip retracted (blue), engaged (green), and scanning (pink) for the frequency ranges (a) 0--200 Hz and (b) 0--800 Hz.}
\end{figure}

\begin{figure}[t]
 \includegraphics[width= 0.5\textwidth]{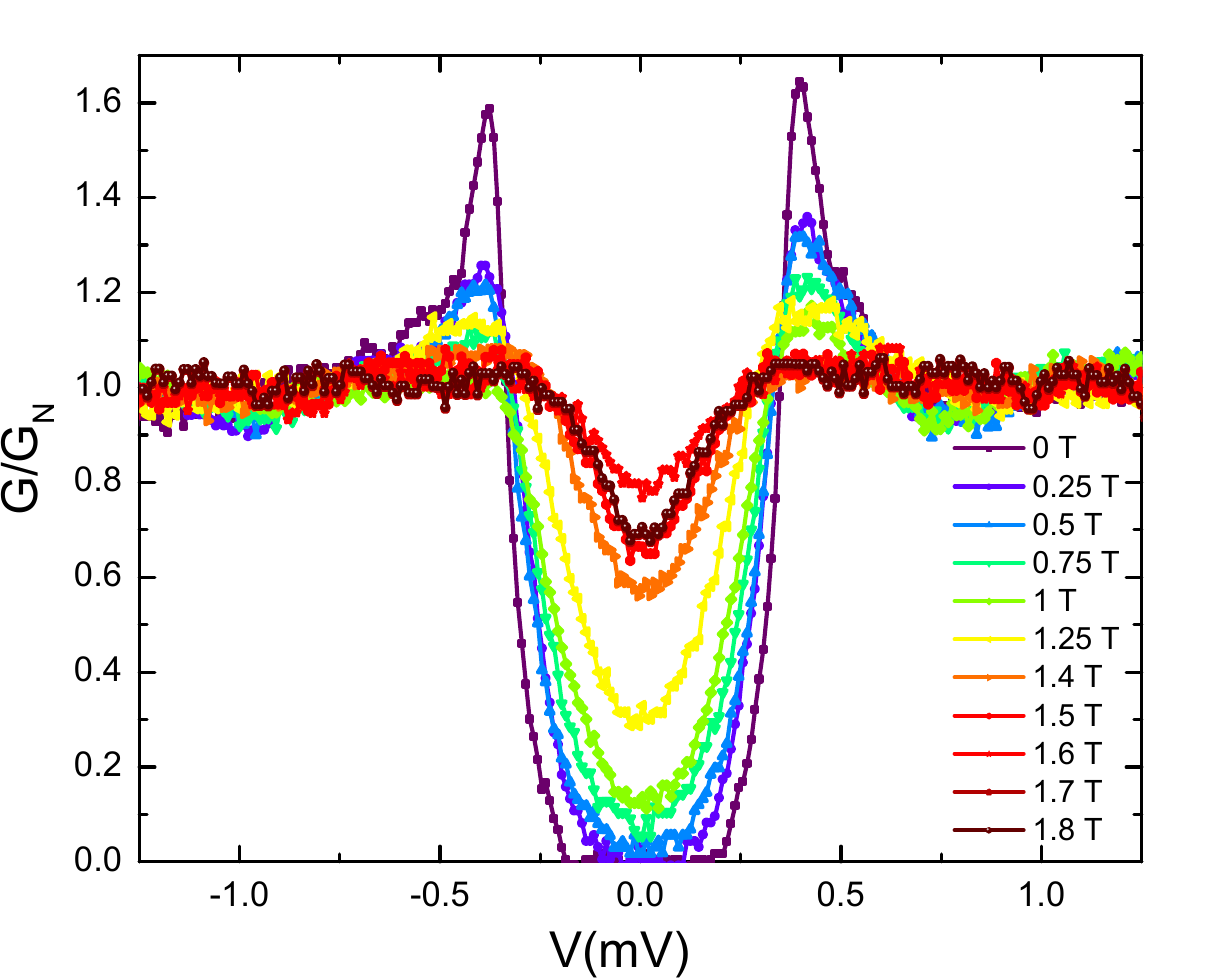}
\caption{(Color Online) Plot of normalized conductance $G/G_n$ versus tip-to-sample voltage $V$ for a Tungsten tip and copper-intercalated Bi$_2$Se$_3$ sample at $T_\text{mix} = 35$ mK. The superconducting gap is noticeably reduced as the magnetic field is increased to 1.8 Tesla.}
\end{figure}
One of the tip-scanner-walker subsystems of the microscope is controlled by an RHK Technology SPM 1000 Scanning Probe Microscope Control System. It has a maximum piezo voltage of 220 V and runs in analog feedback. This subsystem uses an IVP-300 RHK trans-impedance amplifier with a conversion factor of 1 nA/V to maintain the tunnel current. The other tip-scanner-walker subsystem is controlled by a Thermomicroscopes SPM control system (now obsolete), with a digital feedback electronic control unit, and a maximum piezo voltage of 220 V. A variable gain DL Instrument Model 1211 current trans-impedance amplifier monitors the tunnel current. 

\section{Performance}

\begin{figure}[t]
 \includegraphics[width=.5\textwidth]{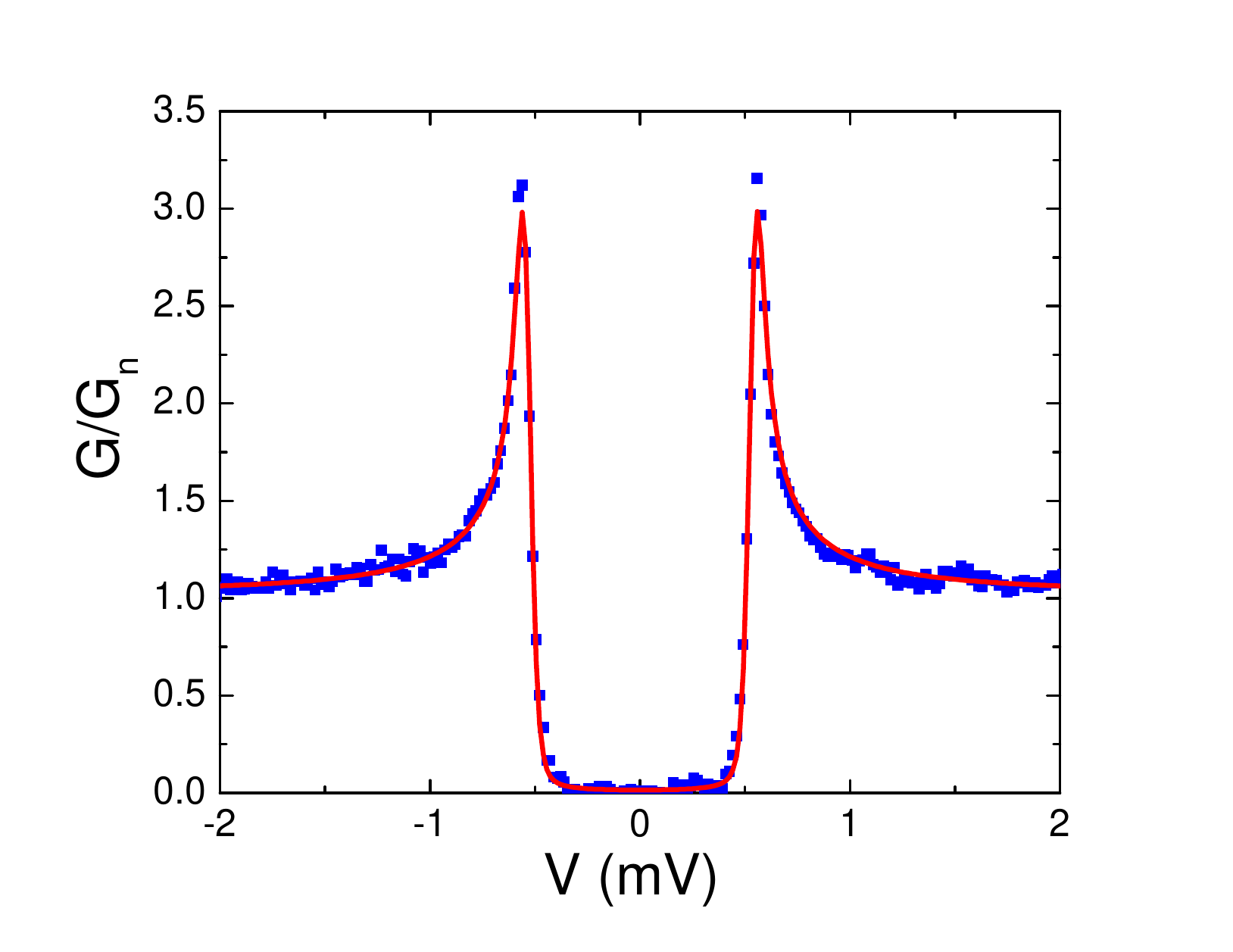}
 \caption{(Color Online) Plot of normalized conductance $G/G_n$ versus tip-to-sample voltage $V$ for a Nb tip and Au sample at $T_\text{mix} = 35$ mK. Blue points are measured data and red curve is fit to Eq.\ (1) with energy gap $\Delta = 0.54$ meV, temperature $T_\text{eff}=184$ mK, and $\Gamma \approx 10^{-5}$ meV.}
\end{figure}

\begin{figure}[t]
 \includegraphics[width=.45\textwidth]{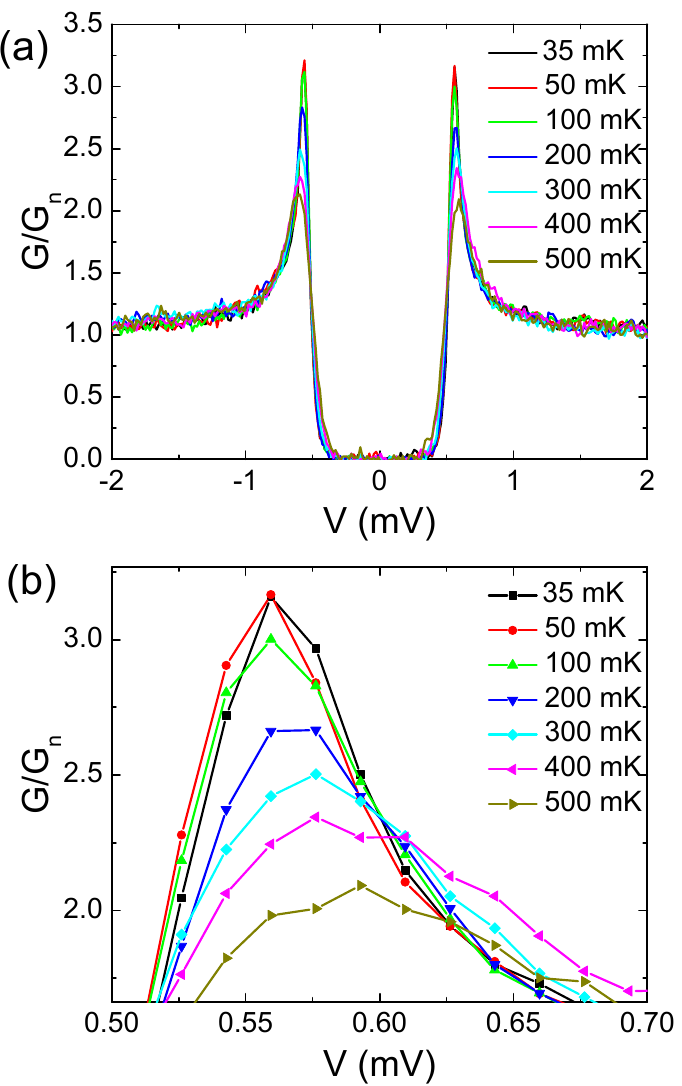}
 \caption{(Color Online) (a) Normalized Conductance G/G$_n$ versus tip voltage $V$ measured  from 35 mK to 500 mK for Nb tip and $\text{Bi}_2\text{Se}_3$ sample. (b) Detailed view of right conductance peaks near 0.6 mV reveals significant temperature dependence to the peak curves.}
\end{figure}

\begin{figure}[t]
 \includegraphics[width=.5\textwidth]{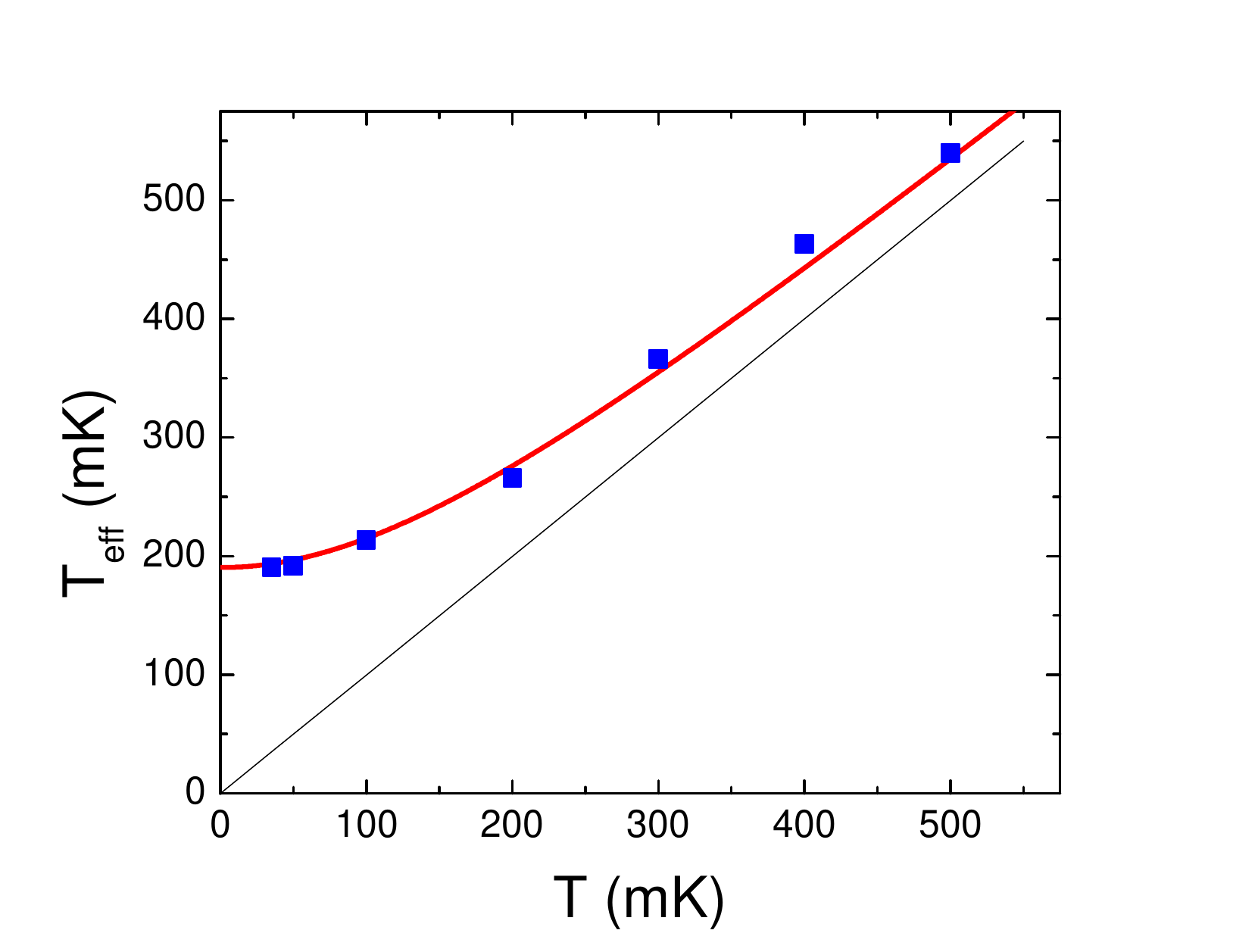}
\caption{(Color Online) Plot of effective temperature $T_\text{eff}$ vs.\ temperature $T$ of the mixing chamber. Blue points were extracted from fitting BCS theory to the data shown in Fig.\ 9. The red curve is Eq.\ (3) with $T_\text{mix} = T$ and $T_0 = 184$ mK. The straight line shows $T_\text{eff} = T$.}
\end{figure}

Figure 5 shows unfiltered atomic resolution images taken with the inner and outer tips on a $\text{Bi}_2\text{Se}_3$ sample. The peak to valley atomic corrugation over the line sections shown of the outer and inner tips are  $\sim$18 pm and $\sim$8 pm respectively. To characterize the noise, we Fourier transformed these images, filtered out the atomic lattice, and then analyzed the inverse Fourier transform.  The rms roughness of the resulting background noise for the outer and inner tip images were $3.55 \pm 0.03$ pm and $1.75 \pm 0.01$ pm, respectively. The noise level for each tip is thus well below the signal for atomic scale features. We note that these measurements were taken with the 1K pot running. We calibrated the scanners in the $x$, $y$, and $z$ directions from these images and images of mono-atomic steps on Au(100). 

Low frequency noise is often the limiting factor in the data obtained by an STM. Figures 6 and 7 show the spectral noise density of the inner and outer tips of our STM while the tip is retracted, in tunneling range, and scanning. The data was taken in typical operating conditions at 30 mK with the 1K pot running. In each case, the electronic noise floor---measured with the tip retracted---is between $10^{-16}$ and $10^{-15}$ A/$\sqrt{\text{Hz}}$. When the tip is engaged, the mechanical noise results in the background level being raised to between $10^{-14}$ and $10^{-13}$ A/$\sqrt{\text{Hz}}$ at low frequencies. Although 60 Hz peaks and higher harmonics are noticeable with both tips, we note that worst the peak signal is $\approx$ 1 pA for the inner tip, and $\approx$ 100 pA for the outer tip. The low  current noise characteristics indicate that the instrument will yield a good signal to noise ratio for both  topographic and spectroscopic data.

The magnet dewar was tested on the system, using a standard single tip Pan-style STM, a Tungsten tip, and a Cu$_x$Bi$_2$Se$_3$ superconducting sample. Fig. \ 8 demonstrates the performance of the magnet via a series of spectroscopy curves measured at different field strengths. The superconducting gap and coherence peaks were predictably diminished with increasing magnetic field.

To obtain the effective temperature and energy resolution of our instrument, conductance ($dI/dV$) measurements were performed between a superconducting tip and a normal sample. The effective temperature can be extracted by fitting the data to a theory that describes quasiparticle tunneling in an SIN tunnel junction \cite{Bardeen61}. Figure 9 shows $dI/dV$ measurements using a Nb inner tip and a $\text{Bi}_2\text{Se}_3$ sample at 35 mK. The measurements were taken by modulating the junction voltage with a small sinusoidal ac voltage and detecting the corresponding response signal with a lock-in technique. Examining the plot, we see that the gap is $\Delta \approx 0.7$ meV, which is about half that of bulk niobium due to finite-size effects. 

 Figure 10(a) shows similar $dI/dV$ curves measured as a function of temperature from 35 mK to 500 mK, and Fig.\ 10(b) shows a closeup of the right conduction peak. We note that although the difference between the 35 mK curve and the 50 mK curve is negligible, there is a noticeable difference between the 50 mK curve and the 100 mK curve, as well as between subsequent curves.  This suggests that our sample was able to cool to temperatures on the order of 100 mK. 

In order to obtain an estimate for the effective temperature of our instrument, we fit each curve in Fig.\ 10 to
\begin{equation}
\frac{dI}{dV} = G_n \int_{-\infty}^{\infty} g(E+eV) N_\text{s,tip}(E)dE,
\end{equation}
where $G_n$ is the normal conductance,  $g(E+eV)= -\partial f(E+eV)/\partial E$, $f(E)$ is the Fermi function at energy $E$, and $N_\text{s,tip}(E)$ is the normalized local density of states of the tip. For a superconducting tip with a BCS density of states and a finite quasiparticle relaxation time $\Gamma$ \cite{Dynes78}, we have: 
  \begin{equation}
N_\text{s,tip}(E) = \text{Re} \left[ \frac{ \vert E- i\Gamma \vert}{\sqrt{(E - i\Gamma)^2 - \Delta^2}}\right] .
\end{equation}
Fitting our data using a weighted least-squares minimization via the Levenberg-Marquardt  \cite{Levenberg44, Marquardt63} algorithm, we extract $\Delta$, $\Gamma$,  and the effective temperature $T_\text{eff}$, which enters through the Fermi function. 

Figure 11 shows a plot of the effective temperature $T_\text{eff}$ versus the temperature $T_\text{mix}$ of the mixing chamber. The solid curve in Fig.\ 11 shows that our data is well explained by a simple model of votage noise \cite{le_Sueur06}
\begin{equation}
T_\text{eff} = \sqrt {T_\text{mix}^2 + T_0^2 },
\end{equation}
where the fitting parameter $T_0 = \sqrt{{3e^2V_n^2}/{k_\text{B}^2\pi^2}}$ is the minimum effective sample temperature due to rms voltage noise $V_n$ (see Appendix A).  From the fit, we find $T_0 =184 \pm 6$ mk. This is equivalent to an energy resolution of $\approx 16$ $\mu$eV. The good agreement between our data and Eqs.\ (1) and (3) suggests that we are limited by voltage noise and that additional cryogenic filtering on the scanner and walker wires would reduce the effective temperature.

\section{Conclusion}

We have presented the design and performance of an STM with two independent tips that can operate at millikelvin temperatures and in high magnetic fields. A connected UHV sample preparation chamber and transfer chamber enable \textit{in situ} sample preparation and transfer to the dilution refrigerator. We have analyzed system performance by characterizing our noise levels and z-stability of both STM tips in atomic resolution images. At a mixing chamber temperature of 35 mK, we estimate the effective temperature of our instrument to be $\approx$ 184 mK, corresponding to an energy resolution of $\approx 16 \; \mu$eV, indicating that phenomena corresponding to extremely low energy scales may be probed using this instrument. 

\section{Acknowledgments}
The authors thank I. Miotkowski and Y.P. Chen for the $\text{Bi}_2 \text{Se}_3$ sample and acknowledge many useful discussions with D. Sullivan and B. Barker. Portions of this work were funded by the NSF under DMR-0605763.

\appendix

\section{}

\subsection*{Spectroscopy with a superconducting tip}

The quasiparticle tunnel current between an STM tip and sample \cite{Bardeen61} is given by
\begin{equation} 
\begin{split}
I(V) = \frac{4\pi e \vert M \vert^2}{\hbar} \int_{-\infty}^{\infty} &\left[ f_\text{tip}(E) - f_\text{sample}(E + eV) \right] \times \\
&\phantom{[} \rho_\text{tip}(E) \rho_\text{sample}(E + eV) dE,
\end{split}
\end{equation}
where $\vert M \vert$ is the average value of the tunneling matrix element assumed to be independent of energy, $f_\text{tip}$ and $f_\text{sample}$ are the electron energy distribution functions in the electrodes which reduce to the Fermi distribution in thermal equilibrium, and $\rho_\text{tip}$ and $\rho_\text{sample}$ are the local electronic density of states (LDOS) of the tip and sample at the point of contact. 

For a normal metal sample, we assume that the density of states of the sample is constant, and can write
\begin{equation} 
I = \frac{G_n}{e \rho_\text{tip,n}(0)} \int_{-\infty}^{\infty} \left[f(E) - f(E+eV)\right] \rho_\text{tip}(E) dE,
\end{equation}
where $\rho_\text{tip,n}(0)$ is the density of states of the tip at the Fermi energy when the tip is in the normal state, and $G_n =  4\pi e^2 \vert M \vert^2 \rho_\text{sample}(0) \rho_\text{tip,n}(0)/\hbar$ is the tunnel conductance in the normal state. Taking a derivative with respect to $V$, we obtain
\begin{equation} 
\frac{dI}{dV} =  \frac{- G_n}{e}\int_{-\infty}^{\infty} \frac{\partial}{\partial V} f(E+eV) \frac{\rho_\text{tip}(E)}{\rho_\text{tip,n}(0)}dE.
\end{equation}
It is convenient to write $g(E)=-f'(E)$, in which case $g$ will be positive everywhere and may be considered a probability distribution. For a superconducting tip this yields
\begin{equation} 
\label{didv_integral}
\frac{dI}{dV} = G_n \int_{-\infty}^{\infty} g(E'+eV)N_\text{s}(E)dE,
\end{equation}
where the normalized density of states of the tip $N_\text{s}(E)={\rho_\text{tip}(E)}/{\rho_\text{tip,n}(0)}$ is given by \cite{Bardeen57}
\begin{equation} 
N_\text{s}(E) = \text{Re}\left[\frac{ \vert E \vert}{\sqrt{E^2 - \Delta^2}} \right].
\label{BCS}
\end{equation}

Dynes incorporated finite quasiparticle lifetime effects into the density of states by including broadening of the singularity at the gap edge \cite{Dynes78}.  Adding an imaginary term $i\Gamma$ to the energy, Eq.\ \eqref{BCS} becomes
\begin{equation}
N_\text{s}(E) = \text{Re} \left[ \frac{ \vert E- i\Gamma \vert}{\sqrt{(E - i\Gamma)^2 - \Delta^2}}\right]
\end{equation}
The recombination rate $\Gamma/\hbar$ is the rate at which a quasiparticle near the energy gap edge scatters inelastically or recombines into the superfluid condensate.

\subsection*{Voltage noise and effective temperature}

Voltage noise fluctuations can be characterized by a probability distribution function $P(V)$ that we assume to be a Gaussian with zero mean and standard deviation $V_n$. The experimentally observed conductance may then be written as
\begin{equation}
\begin{split}
\left\langle \frac{dI}{dV}(V)\right\rangle &= \int P(V') \frac{dI}{dV}(V+V') dV'  \\
&= G_n \int N_\text{s}(E) \int P(V') g(E+e(V+V')) dV'dE \\
&= G_n \int N_\text{s}(E) (P\star g)(E+eV) dE.
\end{split}
\end{equation}
where we have used Eq.\ \eqref{didv_integral} and used $\star$ to represent the convolution of P and g.  Setting $F = P\star g$, we have
\begin{equation}
\left\langle \frac{dI}{dV} \right\rangle \propto \int N_\text{s,tip}(E) F(E) dE.
\end{equation}
We note that $F(E)$ is sharply peaked with standard deviation $\sigma_F$  and determines the sharpness of features such as the increase in conductance at the gap.

Because $F$ is a convolution of $P$ and $g$, which may both be considered probability distributions with mean zero, their variances add and we can write 
\begin{equation}
\sigma_F^2 = e^2\sigma_P^2+\sigma_g^2.
\end{equation}
where the factor of $e$ comes from the convolution in Eq.\ (A7). For a Fermi distribution it can be shown that $\sigma_g^2 = \int^\infty_{-\infty} E^2 g(E) dE = k_\text{B}^2 T^2 \pi^2 / 3$.  Since $N_\text{s}$ is also a sharply peaked function, the total width of the measured coherence peak, $\vert \sigma_{dI/dV} \vert$, should scale as
\begin{equation} 
\sigma_{dI/dV} \propto \sigma_g^2 + e^2\sigma_P^2+\sigma_{N_\text{s}}^2
\end{equation}
where the standard deviation of the voltage noise is given by $\sigma_P = V_n$ and $\sigma_{N_\text{s}} = \Gamma$ is a measure of the width of $\rho$. This yields
\begin{equation} 
\label{noise}
\sigma_{dI/dV} \propto \left( \frac{k_\text{B}^2 T^2 \pi^2}{3} + e^2V_n^2 + \Gamma^2 \right)^{1/2} 
\end{equation}
From Eq.\ \eqref{noise}, we see that the effect of voltage noise is indistinguishable from that of excess temperature when it comes to the broadening of $I$-$V$ characteristics. Assuming $\Gamma^2 \ll e^2 V_n^2$, and setting $k_\text{B}^2 T_\text{eff}^2 \pi^2/3 = k_\text{B}^2 T^2 \pi^2/3 +e^2V_n^2$, we can define the effective temperature of the instrument as \cite{le_Sueur06}:
\begin{equation} 
T_\text{eff} = \sqrt {T^2 + \frac{3 e^2 V_n^2}{k_\text{B}^2 \pi^2} } 
\end{equation}

\newpage

\end{document}